\documentclass[12pt, pra, groupedaddress, amsmath, amssymb, nofootinbib, superscriptaddress]{revtex4-2}
\usepackage{graphicx}
\usepackage{dcolumn}
\usepackage{mathrsfs}
\usepackage{amssymb}
\usepackage{amsthm}
\usepackage{svg}
\usepackage{bm}
\usepackage{bbm}
\usepackage{hyperref}
\usepackage{subcaption}
\usepackage{mathtools}
\usepackage{multirow}
\usepackage{lineno}
\usepackage{amsthm}
\newtheorem*{theorem}{Theorem}
\usepackage[dvipsnames]{xcolor}
\hypersetup{
 colorlinks=true,
 citecolor=blue,
 linkcolor=red,
 urlcolor=blue,
 pdfpagemode=UseNone,
 pdfstartview=FitH}
\usepackage{lipsum}
\usepackage{physics}
\newcommand{\mycomment}[1]{}

\begin{document}

\raggedbottom

\title{Coupling effect of nearest-neighbor interacting qubit chains to a single qubit system}


\author{Manuel Allan L. Orongan}
\affiliation{Department of Physics, School of Foundational Studies and Education, Mapúa University, Intramuros, Manila, Philippines}
\author{Lemuel John F. Sese}
\affiliation{Department of Physics, Pohang University of Science and Technology, Pohang, South Korea}
\author{Rayda P. Gammag}
\email{rpgammag@mapua.edu.ph}
\affiliation{Department of Physics, School of Foundational Studies and Education, Mapúa University, Intramuros, Manila, Philippines}

\begin{abstract}
This study establishes a theorem that provides sufficient criteria for identifying terms in any time-independent Hamiltonian that have no influence on local dynamics, local observables, or any local phenomena, without any approximation. The usefulness of this theorem is demonstrated by predicting the behavior of three systems. The first system consists of multiple qubit chains with Ising interactions. The second system is formed by a single qubit chain, differentiated by the substitution of a Dzyaloshinskii-Moriya (DM) interaction for the last Ising interaction. The predictions were verified by analytically deriving the reduced dynamics of both systems. A third system was also considered, namely, a short qubit chain with a transverse magnetic field on the intermediary environment qubit. This transverse magnetic field signals that the commutation relation required by our theorem no longer holds. The third system's local dynamics were also derived analytically, demonstrating that all Hamiltonian constituents contributed to the reduced dynamics. The physical consequences of our theorem’s inapplicability were also explored using this third system by analyzing the entanglement dynamics between the qubits that do not directly interact. The results showed that the entanglement is caused by an \textit{emergent coupling} between the two non-directly interacting subsystems. This \textit{emergent coupling} arises from the non-commutativity of the intermediate interactions connecting the two subsystems. When searching for these \textit{emergent couplings}, our theorem can eliminate intermediate interactions that will not produce them.
\end{abstract}


\maketitle
\section{Introduction}

Quantum mechanics is our most fundamental description of physical phenomena, with closed quantum systems being an idealization of real systems. This is because no physical system can be truly isolated from the rest of the universe. As such, the theory of open quantum systems was developed to analyze the behavior of quantum systems interacting with an environment \cite{Breuer}. The general setup is shown in figure \ref{fig:SystemT}, where we have a closed total system with a subsystem serving as the local system and the rest treated as its environment. It is common to see states in open quantum systems expressed as density matrices $\hat{\rho}$ rather than the more familiar state vector $\ket{\Psi}$. This is useful as interactions with an environment cause locally non-unitary time-evolution. Thus, the equivalent pure state density matrix of any state vector, i.e. $\hat{\rho} = \ketbra{\Psi}$, inevitably becomes a mixed state, which has no state vector counterpart \cite{Manzano_2020}. The framework of open quantum systems continues to deepen our understanding of condensed matter \cite{Manzano_2016, Chen_2024}, quantum biology \cite{Plenio_2008, Manzano_2013, Ruan_2026}, quantum information \cite{Tittel_2000, Bose_2003, Marchukov_2016}, atomic physics \cite{Mao_2025, Alporha_2025}, and the quantum-to-classical transition \cite{Sese_2022, Cucchietti_2005, Tessieri_2002, Galapon_2016, Mirkin_2021, Marchisio_2026}.

Many studies simplify their models by preventing entities within their environment from interacting with each other \cite{Zurek_1982, Cucchietti_2005, Sese_2022, Alporha_2025, Gedik_2006}. However, it is more realistic to allow environment components to interact with one another. As such, many studies have also analyzed local phenomena subject to environments with intra-environment interactions. These interactions would take on various forms, such as Heisenberg interactions \cite{Bose_2003, Marchukov_2016}, Dzyaloshinskii-Moriya (DM) interactions \cite{Tchoffo_2016, Mahmoudi_2020}, and three-spin interactions \cite{Mahmoudi_2017, Mahmoudi_2020}, among others. These studies discovered significant contributions from these interactions, particularly on the transfer of information and the formation of entanglement among adjacent qubits. However, a glaring inconsistency arises when applying these results to the system considered by Dannenberg \cite{Dannenberg_2008}. Within his system, he observed that local dynamics were identical whether or not there are intra-environment interactions.

We are motivated by Dannenberg's observation to produce a theorem that provides sufficient criteria for determining when certain interactions within the environment become irrelevant to local dynamics. With this theorem, we derive the dynamics of three systems. The first two are systems that can be simplified by our theorem, while the last system cannot. We also further explore the consequences of our theorem's inapplicability in this last system by analyzing the entanglement dynamics between non-adjacent qubits.

\section{The Theorem}\label{Theorem}

We state the theorem being introduced by this study along with its mathematical proof.

\begin{theorem}
    Let $\hat{H}$ be a time-independent Hamiltonian that can be broken apart into the local system Hamiltonian $\hat{H}_{\mathcal{S}}$, the interaction Hamiltonian $\hat{H}_\text{int}$, and the environment Hamiltonian $\hat{H}_\mathcal{E}$. If $\hat{H}_\mathcal{E}$ can be further subdivided into $\hat{H}_{\mathcal{E}_1}$ and $\hat{H}_{\mathcal{E}_2}$ such that $\hat{H}_{\mathcal{E}_2}$ encompasses all terms of $\hat{H}_\mathcal{E}$ that maintain the commutation relation $[\hat{H}, \hat{H}_{\mathcal{E}_2}]=0$, then terms classified under $\hat{H}_{\mathcal{E}_2}$ will not influence local dynamics or any local observables of the system.
    
\end{theorem}

\begin{figure}
    \centering
    \includegraphics[width=0.4\linewidth]{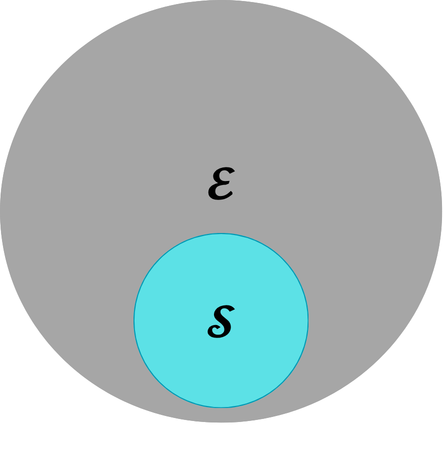}
    \caption{An open quantum system with the local system $\mathcal{S}$ and environment $\mathcal{E}$.}
    \label{fig:SystemT}
\end{figure}

\begin{proof}
    Consider a general open quantum system, as shown in figure \ref{fig:SystemT}, occupying the composite Hilbert space $\mathcal{H}$ of the system $\mathcal{S}$ and the environment $\mathcal{E}$,
    \begin{equation} \label{eq:General Hilbert Space}
        \mathcal{H} = \mathcal{S} \otimes \mathcal{E}.
    \end{equation}
    
    Given the commutation of $\hat{H}_{\mathcal{E}_2}$ with the remaining constituents of $\hat{H}$, there must exist some shared eigenbasis between them $\{ \ket{\psi_m} \} \otimes \{ \ket{\phi_h} \}$ with $m, h \in \mathbbm{N}$ \cite{Sakurai}. Here, $\ket{\psi_m}$ and $\ket{\phi_h}$ are the eigenbasis of $\mathcal{S}$ and $\mathcal{E}$, respectively. Thus, we can express this Hamiltonian in terms of its eigenenergies in this shared basis as
    \begin{align} \label{eqs:Proof Eigenenergies}
        \hat{H}_{\mathcal{E}_2} &= \sum_{m, h} E_{h} \ketbra{\psi_m}{\psi_m} \otimes \ketbra{\phi_h}{\phi_h} \text{ and} \\
        \hat{H} - \hat{H}_{\mathcal{E}_2} &= \sum_{m, h} J_{mh} \ketbra{\psi_m}{\psi_m} \otimes \ketbra{\phi_h}{\phi_h}.
    \end{align}

    The von Neumann equation dictates that the density matrix representing the total system would take the form
    \begin{equation} \label{eq:Proof Density Matrix}
        \hat{\rho} (t) = \sum_{m, n, h, k} \rho_{m n h k} e^{-i \left( J_{mh} - J_{nk} + E_h - E_k \right) t / \hbar} \ketbra{\psi_m}{\psi_n} \otimes \ketbra{\phi_h}{\phi_k},
    \end{equation}
    where the constants $\rho_{mnhk}$ reveal the initial state of the system.

    We partial trace this density matrix over the degrees of freedom of the environment $\{ \ket{\phi_l} \}$. This reveals that the eigenenergies from $\hat{H}_{\mathcal{E}_2}$, namely $E_h - E_k$, cancel out for all surviving terms of the density matrix after partial tracing\footnote{Further discussion about partial tracing can be found in Appendix \ref{ap:Partial Trace}.}. This leaves us with a reduced density matrix of
    \begin{equation} \label{eq:Proof Reduced Density Matrix}
        \hat{\rho}_0 (t)
            = \sum_{l, m, n} \rho_{mnll} e^{-i(J_{ml}-J_{nl})t/\hbar}  \ketbra{\psi_m} {\psi_n}.
    \end{equation}

    We clearly see from equation \eqref{eq:Proof Reduced Density Matrix} that the local dynamics of the system is independent of the eigenenergies described by $\hat{H}_{\mathcal{E}_2}$. We can also prove that the same is true of all local observables of the system. This follows from their expectation value, $\left\langle \hat{O} (t) \right\rangle = \text{Tr}\left[ \hat{O} \, \hat{\rho}_0 (t)\right]$, being only dependent on the state of the local system, which we have already proven is independent of $\hat{H}_{\mathcal{E}_2}$. In fact, the resulting equations describing the dynamics of the local system and of any local observables would be the same had we omitted $\hat{H}_{\mathcal{E}_2}$ from the Hamiltonian.

    With this proof, we can sufficiently claim that terms classified under $\hat{H}_{\mathcal{E}_2}$ will have no influence on the local system. Thus when analyzing only the local system, one can omit these terms and still achieve the same result.
\end{proof}

\section{Sample Hamiltonians}
In this section, we present three sample Hamiltonians: the first one closely resembles the one used by Dannenberg \cite{Dannenberg_2008}, the second incorporates a DM interaction, and the last one cannot be simplified by our theorem. We demonstrate the validity of the theorem’s predictions by deriving the reduced dynamics of each system and observing the surviving terms from their respective Hamiltonians. We also further explore the third system by examining the entanglement dynamics between qubits that are not directly interacting.

\subsection{First Hamiltonian: Multiple Qubit Chains Connected by a Central Qubit}\label{First Hamiltonian}

\begin{figure}
    \centering
    \includegraphics[width=0.9\linewidth]{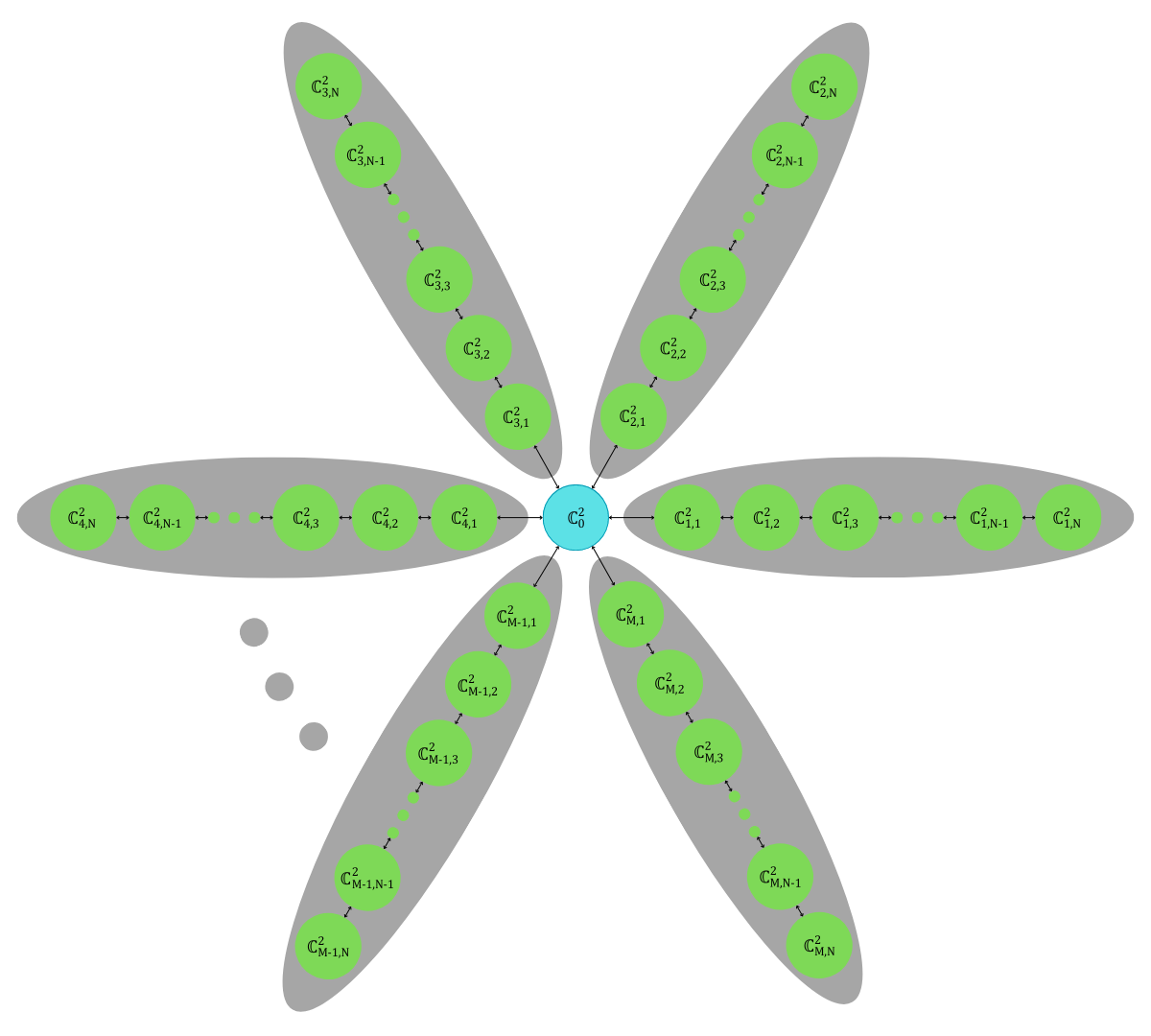}
    \caption{A central qubit (blue) interacting with $M$ Ising chains (gray), each consisting of $N$ qubits (green)}
    \label{fig:System1}
\end{figure}

This first example considers multiple qubit chains connected by a central qubit that interacts with the first qubit of each chain. This is shown in figure \ref{fig:System1}. The central qubit is treated as the local system for which we derive its local dynamics, represented by its reduced density matrix.

This system would have the Hilbert space $\mathcal{H}$ expressed as a tensor product of $\mathbb{C}^2$,
\begin{equation} \label{eq:Hilbert Space 1}
    \mathcal{H} = \mathbb{C}_{0}^{2}
    \bigotimes\limits_{j=1}^{M}
    \bigotimes\limits_{i=1}^{N}
    \mathbb{C}_{j, i}^{2}.
\end{equation}
\noindent
where $\mathbb{C}_{0}^2$ and $\mathbb{C}_{ij}^2$ are the Hilbert spaces of the central qubit and of the qubit on the $i$th site of the $j$th chain, respectively.

The Hamiltonian of this system consists of an external magnetic field acting on the central qubit $H_0$, the Ising interaction between the central qubit and the first qubit of each qubit chain $H_{0\mathcal{E}}$, and the Ising interactions along all chains $H_{\mathcal{EE}}$. Each of these constituents of the total Hamiltonian can be expanded as,
\begin{subequations} \label{eqs:Hamiltonian 1 Expanded}
\begin{align}
    \hat{H}_0 &= \frac{1}{2}\omega_0 \hat{\sigma}_0^z,\\
    \hat{H}_{0\mathcal{E}} &= 
        \frac{1}{2}\hat{\sigma}_{0}^z \sum_{j=1}^M G_{01}^{(j)}
        \hat{\sigma}_{j, 1}^z,\ \text{and} \\
    \hat{H}_{\mathcal{EE}} &= 
        \frac{1}{2}\sum_{j=1}^M \sum_{i=1}^{N-1} G_{i, i+1}^{(j)}
        \hat{\sigma}_{j i}^z \hat{\sigma}_{j, i+1}^z.
\end{align}
\end{subequations}
The operator $\hat{\sigma}^\alpha$ where $\alpha \in \{ x, y, z \}$ are Pauli spin operators with subscripts matching the Hilbert space in which they reside. Meanwhile, the constants $\omega_0, G_{i, i+1}^{(j)} \in \mathbb{R}$ respectively represent the strengths of the external magnetic field acting on the local system and each corresponding interaction, all written in natural units, i.e., $\hbar=1$. Moving forward, we continue to observe the use of natural units.

This Hamiltonian closely resembles that considered by Dannenberg \cite{Dannenberg_2008}, with all interactions being diagonal in the $z$-basis. However, his model considered each qubit interacting with every other qubit instead of arranging them as qubit chains. It is for his system that he observed no influence of intra-environment interactions on local dynamics. Likewise for this system, our theorem predicts that all terms in $\hat{H}_\mathcal{EE}$ have no influence on local dynamics for we can map the Hamiltonian to the general Hamiltonian in our theorem as
\begin{subequations} \label{eqs:Theorem Mapping 1}
\begin{align}
    \hat{H}_\mathcal{S} &= \hat{H}_0, \\
    \hat{H}_{\mathcal{SE}} &= \hat{H}_{0\mathcal{E}}, \\
    \hat{H}_{\mathcal{E}_1} &= 0, \text{ and} \\
    \hat{H}_{\mathcal{E}_2} &= H_{\mathcal{EE}}.
\end{align}
\end{subequations}

We verify this prediction by deriving the reduced density matrix the central qubit.

We first assume that the system is initially separable with the form
\begin{equation} \label{eq:Initial State 1}
    \ket{\Psi(0)} = \left(a_{0}^{(+1)}\ket{+1_{0}}
        + a_{0}^{(-1)}\ket{-1_{0}}\right) \bigotimes\limits_{h=1}^M \bigotimes\limits_{k=1}^N
        \left(a_{hk}^{(+1)} \ket{+1_{hk}}
        + a_{hk}^{(-1)} \ket{-1_{hk}}\right).
\end{equation}
where $\ket{\alpha_0}$ and $\ket{\alpha_{hk}}$ represent the up ($\alpha = +1$) and down ($\alpha = -1$) states of the $z$-basis with subscripts matching the Hilbert space within which they belong. The coefficients of each state are normalized, i.e., $\left|a_0^{(+1)}\right|^2 + \left|a_0^{(-1)}\right|^2=1$ and $\left|a_{hk}^{(+1)}\right|^2 + \left|a_{hk}^{(-1)}\right|^2=1$.

Taking advantage of the commuting nature of all the constituents of the Hamiltonian, we can analyze them separately and note the effect of each term on the time evolution of the reduced density matrix. These analyses reveal that $\hat{H}_\mathcal{S}$ contributes a complex exponential term $e^{\mp i \omega_0 t}$ to the off-diagonal terms $\ket{\pm1_0}\bra{\mp1_0}$ of the reduced density matrix. $\hat{H}_{\mathcal{SE}}$ contributes a product term $\prod\limits_{h=1}^M \left( \left|a_{h, 1}^{(+1)}\right|^2 e^{\mp iG_{01}^{(h)}t} + \left|a_{h,1}^{(-1)}\right|^2 e^{\pm iG_{01}^{(h)}t} \right)$ to those same off-diagonal elements. Lastly, $\hat{H}_\mathcal{EE}$ has no effect on the reduced density matrix. The absence of a contribution from $\hat{H}_\mathcal{EE}$ already demonstrates that these terms do not influence the central qubit's local dynamics. Nonetheless, we still show the final form of the reduced density matrix,
\begin{multline} \label{eq:Reduced Density Matrix 1}
    \hat{\rho}_{0}(t)
        = \left|a_{0}^{(+1)}\right|^2 \ketbra{+1_0}{+1_0}
        + \left|a_{0}^{(-1)}\right|^2 \ketbra{-1_0} {+1_0} \\
        \quad + a_{0}^{(+1)} \left(a_{0}^{(-1)}\right)^*
        e^{-i \omega_0 t}
        \prod\limits_{h=1}^M
            \left( \left|a_{h, 1}^{(+1)}\right|^2 e^{-iG_{01}^{(h)}t}
            + \left|a_{h,1}^{(-1)}\right|^2 e^{iG_{01}^{(h)}t} \right)  \ketbra{+1_0}{-1_0} +\text{h.c.}
\end{multline}

As predicted, clearly no terms within $\hat{H}_\mathcal{EE}$ are present in the reduced density matrix. This implies that such terms do not influence local dynamics, any local observables, or any local phenomena emergent from these. It is worth noting that the reduced density matrix is the same as that shown by Zurek \cite{Zurek_1982}, despite the presence of qubit chains. This same observation was made by Dannenberg for his system \cite{Dannenberg_2008}.

\subsection{Second Hamiltonian: Single Qubit Chain with a Dzyaloshinskii-Moriya Interaction} \label{Second Hamiltonian}

\begin{figure}
    \centering
    \includegraphics[width=0.9\linewidth]{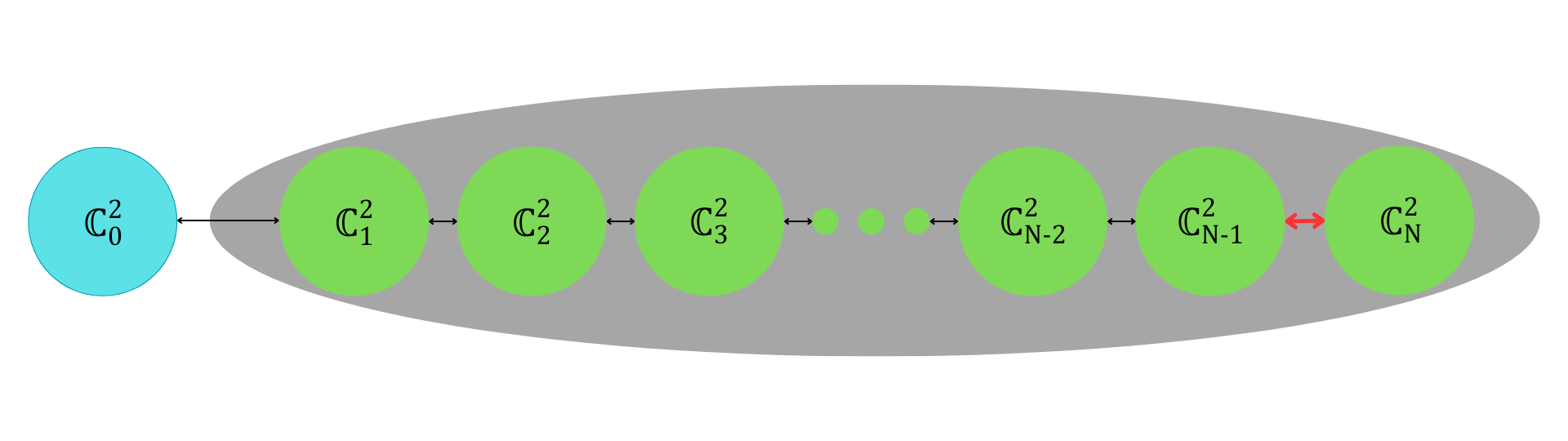}
    \caption{A central qubit (blue) interacting with the first qubit of an $N$ qubit Ising chain (gray) with a single Dzyaloshinskii-Moriya intra-environment interaction (red)}
    \label{fig:System2}
\end{figure}

In this next system, we consider only a single qubit chain as we notice that the effect of multiple chains on local dynamics is merely a product of each chain's individual contribution. Given that our previous example used only interactions that are diagonal in the $z$-basis, we differentiate this next example by making the last interaction contain non-diagonal elements. As such, we use the DM interaction as it is an experimentally observed interaction that exhibits non-diagonal elements \cite{Sergienko_2006, Heide_2008, Belmeguenai_2015}. Again, the local system is a qubit interacting only with the first qubit of the chain, as shown in figure \ref{fig:System2}.

This system would occupy the Hilbert space
\begin{equation} \label{eq:Hilbert Space 2}
    \mathcal{H} = \mathbb{C}_{0}^{2}
    \bigotimes\limits_{i=1}^{N}
    \mathbb{C}_{i}^{2}.
\end{equation}
The definitions for $\mathbb{C}_{0}^{2}$ and $\mathbb{C}_{i}^{2}$ are the same as in equation \eqref{eq:Hilbert Space 1} but without the first subscript for $\mathbb{C}_{i}^{2}$ as there is only one qubit chain.

Similar to the first system, this Hamiltonian consists of an external magnetic field acting on the central qubit $\hat{H}_0$, the Ising interaction between the central qubit and the first qubit of the qubit chain $\hat{H}_{0\mathcal{E}}$, the Ising interactions along the chain $\hat{H}_{\mathcal{EE}}$, and the DM interaction between the last two qubits $\hat{H}_{\text{DM}}$. This DM interaction has a constant strength $D$ and has a DM vector oriented parallel to the action of the Ising interactions. Each of these constituents of $\hat{H}$ can be expanded as
\begin{subequations} \label{eq:Hamiltonian 2 Expanded}
\begin{align}
    \hat{H}_0 &= \frac{1}{2}\omega_0 \hat{\sigma}_0^z, \\
    \hat{H}_{0\mathcal{E}} &= 
        \frac{1}{2} G_{01}
        \hat{\sigma}_{0}^z
        \hat{\sigma}_{1}^z, \\
    \hat{H}_{\mathcal{EE}} &= 
        \frac{1}{2} \sum_{i=1}^{N-2} G_{i, i+1}
        \hat{\sigma}_{i}^z \hat{\sigma}_{i+1}^z, \text{ and}\\
    \hat{H}_{\text{DM}} &=
        \frac{1}{2} D
        \left(
            \hat{\sigma}_{N-1}^x\hat{\sigma}_{N}^y
            - \hat{\sigma}_{N-1}^y\hat{\sigma}_{N}^x
        \right).
\end{align}
\end{subequations}

 Like the first sample Hamiltonian, we can map the constituents of this Hamiltonian to our theorem as
 \begin{subequations} \label{eqs:Theorem Mapping 2}
\begin{align}
    \hat{H}_\mathcal{S} &= \hat{H}_0, \\
    \hat{H}_{\mathcal{SE}_1} &= \hat{H}_{0\mathcal{E}}, \\
    \hat{H}_{\mathcal{E}_1} &= 0, \text{ and} \\
    \hat{H}_{\mathcal{E}_2} &= \hat{H}_{\mathcal{EE}} + \hat{H}_\text{DM}.
\end{align}
\end{subequations}
Once again, the theorem predicts that local dynamics are only influenced by $\hat{H}_0$ and $\hat{H}_{0\mathcal{E}}$, which we can again verify by deriving the reduced density matrix of the local system.

Again, we assume the total system is initially separable, i.e.,
\begin{equation} \label{eq:Initial State 2}
    \ket{\Psi(0)} = \left(a_{0}^{(+1)}\ket{+1_{0}}
        + a_{0}^{(-1)}\ket{-1_{0}}\right) \bigotimes\limits_{k=1}^N
        \left(a_{k}^{(+1)} \ket{+1_{k}}
        + a_{k}^{(-1)} \ket{-1_{k}}\right).
\end{equation}

From our first example, we already know the dynamics of this system without the DM interaction. Thus, we only need to find the contribution of this unique term. However, the DM interaction's non-commuting nature with the adjacent Ising interaction between qubits $N-2$ and $N-1$ makes the trivial decomposition we utilized in the earlier example no longer applicable. Instead, we employ the interaction picture\footnote{Further discussion of the interaction picture can be found in Appendix \ref{ap:Interaction Representation} along with the differential equations that give rise to the interaction picture state vectors in equations \eqref{eq:Interaction State Vector 2} and \eqref{eq:Interaction State Vector 3} with their respective coefficients.} by treating the DM interaction as the \textit{interaction part} of the Hamiltonian. Thus, the interaction potential for this system is
\begin{align}
    \hat{H}^{(\text{I})}_{\text{DM}}(t) &= e^{i (\hat{H} - \hat{H}_{\text{DM}}) t}
        \hat{H}_{\text{DM}}
        e^{-i (\hat{H} - \hat{H}_{\text{DM}}) t} \label{eq:Interaction Potential 2 General} \\
        & = \frac{1}{2}D
            e^{
                -i G_{N-2, N-1}
                    \hat{\sigma}_{N-2}^z \hat{\sigma}_{N-1}^z t
            }
            (\hat{\sigma}_{N-1}^x\hat{\sigma}_{N}^y
                - \hat{\sigma}_{N-1}^y\hat{\sigma}_{N}^x),
    \label{eq:Interaction Potential 2 Solved}
\end{align}
where the superscript $(\text{I})$ distinguishes it as the interaction representation.

We solve the equations of motion of the interaction picture to get the state vector's interaction representation,
\begin{multline}
    \ket{\Psi^{(\text{I})}(t)}
        = \bigotimes_{n = 0}^{N-3}
            \left(
                a_{n}^{(+1)} \ket{+1_n}
                + a_{n}^{(-1)} \ket{-1_n}
            \right) \\
            \otimes
                \sum_{\alpha, \beta, \gamma}
            \left(
                a_{N-2}^{(\alpha)} a_{N-1}^{(\beta)} a_{N}^{(\gamma)}
                    e^{
                        i \frac{\sqrt{
                                \left(\overline{G}_{\alpha \beta}\right)^2
                                + \left(\overline{D}_{\beta \gamma}\right)^2
                            }} {2} t
                    }
                + i A_{\alpha \beta \gamma}
                    \operatorname{sin} \left(
                        \frac{\sqrt{
                            \left(\overline{G}_{\alpha \beta}\right)^2
                            + \left(\overline{D}_{\beta \gamma}\right)^2
                        }} {2} t
                    \right)
            \right) \\
                \times e^{
                    i \overline{G}_{\alpha \beta} t/2
                }
        \ket{\alpha_{N-2}}
            \ket{\beta_{N-1}}
                \ket{\gamma_{N}},
    \label{eq:Interaction State Vector 2}
\end{multline}
where $\overline{G}_{\alpha \beta} = \alpha \beta G_{N-2, N-1}$ and $\overline{D}_{\beta \gamma} = (\beta - \gamma)D$. The normalizing coefficient $A_{\alpha \beta \gamma}$ is expressed in terms of the initial probability amplitudes as
\begin{equation} \label{eq:Coefficient 2}
    A_{\alpha\beta\gamma} =
        \frac{
            \left(
                \sqrt{
                        \left(\overline{G}_{\alpha \beta}\right)^2
                        + \left(\overline{D}_{\beta \gamma}\right)^2
                    }
                - \overline{G}_{\alpha \beta}
            \right)
                a_{N-1}^{(\beta)} a_{N}^{(\gamma)}
            - i \overline{D}_{\beta \gamma}
                a_{N-1}^{(-\beta)} a_{N}^{(-\gamma)}
            }
            {
                \sqrt{
                    \left(\overline{G}_{\alpha \beta}\right)^2
                    + \left(\overline{D}_{\beta \gamma}\right)^2
                }
            }
                a_{N-1}^{(\alpha)}.
\end{equation}

We now get the effects of the DM interaction on local dynamics by taking the reduced density matrix of the state vector in the interaction picture,
\begin{equation} \label{eq:Interaction Picture Reduced Density Matrix 2}
    \hat{\rho}_0^{(\text{I})}{(t)}
        =\left|a_{0}^{(+1)}\right|^2 \ketbra{+1_0}
        + \left|a_{0}^{(-1)}\right|^2 \ketbra{-1_0}
        + a_{0}^{(+1)} \left(a_{0}^{(-1)}\right)^* \ketbra{+1_0}{-1_0}
        + \text{h.c.}
\end{equation}

We see that the DM interaction has no effect on the time-evolution of the local system. Combining this observation with the contribution from the other constituents of this Hamiltonian as observed in the previous example, we get the reduced density matrix for this system
\begin{multline} \label{eq:Reduced Density Matrix 2}
    \hat{\rho}_{0}(t)
        = \left|a_{0}^{(+1)}\right|^2 \ketbra{+1_0}
        + \left|a_{0}^{(-1)}\right|^2 \ketbra{-1_0} \\
        \quad + a_{0}^{(+1)} \left(a_{0}^{(-1)}\right)^*
        e^{-i \omega_0 t}
        \left( \left|a_{1}^{(+1)}\right|^2
            e^{-iG_{0,1}t}
            + \left|a_{1}^{(-1)}\right|^2 e^{iG_{0,1}t} \right)  \ketbra{+1_0}{-1_0} + \text{h.c.}
\end{multline}
Just as our theorem had predicted, the local dynamics of this system are only governed by $\hat{H}_0$ and $\hat{H}_{0\mathcal{E}}$ despite the inclusion of a unique DM interaction.

\subsection{Third Hamiltonian: Short Qubit Chain with a Transverse Magnetic Field}\label{Third Hamiltonian}

For this last sample Hamiltonian, we explore an example for which our theorem is not applicable. For the sake of simplicity, we maintain the environment consisting of a single qubit chain but shorten it to two qubits with the local qubit still interacting only with the first qubit of the chain. Instead of adding a DM interaction, we apply a transverse magnetic field to the first environment qubit. This breaks the commutation relation needed by our theorem to classify the intra-environment interaction as part of $\hat{H}_{\mathcal{E}_2}$, thus preventing us from claiming that this interaction has no influence on the local state.

As before, we start by defining this system's Hilbert space as
\begin{equation} \label{eq:Hilbert Space 3}
    \mathcal{H} = \mathbb{C}_0^2 \otimes \mathbb{C}_1^2 \otimes \mathbb{C}_2^2
\end{equation}
and its Hamiltonian consisting of an external magnetic field applied to the central qubit $\hat{H}_0$, the interaction between the central qubit and the first qubit of the chain $\hat{H}_{0\mathcal{E}}$, and the interaction between the qubits within the chain $\hat{H}_\mathcal{EE}$ with the inclusion of an external magnetic field $\hat{H}_1$ transverse to the orientation of the Ising interactions. We can express the constituents of this Hamiltonian as
\begin{subequations} \label{eqs:Hamiltonian 3 Expanded}
\begin{align}
    \hat{H}_0 &= 
        \frac{1}{2} \omega_0 \hat{\sigma}_0^z, \\
    \hat{H}_{0\mathcal{E}} &= 
        \frac{1}{2} G_{01} \hat{\sigma}_{0}^z
        \hat{\sigma}_{1}^z, \\
    \hat{H}_1 &=
        \frac{1}{2} \omega_1 \hat{\sigma}_1^x, \text{ and} \\
    \hat{H}_{\mathcal{EE}} &= 
        \frac{1}{2} G_{12} \hat{\sigma}_{1}^z
        \hat{\sigma}_{2}^z.
\end{align}
\end{subequations}

Clearly, no constituent of this Hamiltonian satisfies the commutation relations required by our theorem to be classified under $\hat{H}_{\mathcal{E}_2}$. Thus, the theorem cannot isolate any terms that have no influence on local dynamics.

We continue by demonstrating that all terms of this Hamiltonian participate in influencing local dynamics by deriving its reduced density matrix.

As before, we assume a separable initial state,
\begin{equation} \label{eq:Initial State 3}
    \ket{\Psi(0)} = \left(a_{0}^{(+1)}\ket{+1_{0}}
        + a_{0}^{(-1)}\ket{-1_{0}}\right) \bigotimes\limits_{k=1}^2
        \left(a_{k}^{(+1)} \ket{+1_{k}}
        + a_{k}^{(-1)} \ket{-1_{k}}\right).
\end{equation}

Excluding $\hat{H}_1$, this Hamiltonian is identical to that analyzed in the first example if we were to consider only a single chain with a length of two qubits. As such, we only need to find the contribution of $\hat{H}_1$ to the reduced density matrix. Similar to the second sample Hamiltonian's DM interaction, the non-commuting nature of $\hat{H}_1$ with the other constituents of this Hamiltonian means we cannot utilize the trivial decomposition used in the first example; instead, we once again employ the interaction picture, setting this unique term as the \textit{interaction part} of the Hamiltonian. Thus, the interaction potential is
\begin{align}
    \hat{H}^{(\text{I})}_1(t) =
    \frac{1}{2} \omega_1
        e^{
            i \left(
                G_{01} \hat{\sigma}_0^z \hat{\sigma}_1^z
                + G_{12} \hat{\sigma}_1^z \hat{\sigma}_2^z
            \right) t
        } \hat{\sigma}_1^x.
    \label{eq:Interaction Potential 3 Solved}
\end{align}

This interaction potential results in the state vector
\begin{multline} \label{eq:Interaction State Vector 3}
    \ket{\Psi^{(\text{I})}(t)} =
        \sum_{\alpha, \beta, \gamma}
            \left[
                a_0^{(\alpha)} a_1^{(\beta)} a_2^{(\gamma)}
                    e^{
                        i \frac{\sqrt{
                                \omega_1^2
                                + \left(\overline{G}_{\alpha \beta \gamma}\right)^2
                            }} {2} t
                    }
                + i A_{\alpha \beta \gamma}
                    \operatorname{sin} \left(
                        \frac{\sqrt{
                            \omega_1^2
                            + \left(\overline{G}_{\alpha \beta \gamma}\right)^2
                        }} {2} t
                    \right)
            \right] \\
                \times e^{
                    i \overline{G}_{\alpha \beta \gamma} t/2
                }
        \ket{\alpha_0} \ket{\beta_1} \ket{\gamma_2},
\end{multline}
where $\overline{G}_{\alpha \beta \gamma} = \beta \left( \alpha G_{0, 1} + \gamma G_{1, 2} \right)$. Similar to equation \eqref{eq:Interaction State Vector 2}, $A_{\alpha \beta \gamma}$ is a normalizing coefficient given in terms of the initial probability amplitudes as
\begin{equation} \label{eq:Coefficient 3}
    A_{\alpha \beta \gamma} =
        \frac{
            \left(
                \sqrt{
                    \omega_1^2
                    + \left(\overline{G}_{\alpha \beta \gamma}\right)^2
                }
                - \overline{G}_{\alpha \beta \gamma}
            \right) a_1^{(\beta)}
            - \omega_1 a_1^{(-\beta)}
        }
        {
        \sqrt{
            \omega_1^2
            + \left(\overline{G}_{\alpha \beta \gamma}\right)^2}
        } a_0^{(\alpha)} a_2^{(\gamma)}.
\end{equation}

Converting this state vector to its Schr{\"o}dinger picture counterpart and partial tracing the equivalent density matrix reveals the reduced density matrix as
\begin{multline}
    \hat{\rho}_0(t) =
        \left| a_0^{(+1)} \right|^2
            \ketbra{+1_0}
        + \left| a_0^{(-1)} \right|^2
        \ketbra{-1_0} \\
        + \sum_{\beta, \gamma}
        e^{-i \omega_0 t} \left( \left[
            a_0^{(+1)} a_1^{(\beta)} a_2^{(\gamma)}
                e^{
                    -i \frac{\sqrt{
                        \omega_1^2
                        + \overline{G}_{+1, \beta \gamma}^2
                    }}{2} t
                }
            + i A_{+1, \beta \gamma}
                \operatorname{sin}\left(
                    \frac{\sqrt{
                        \omega_1^2
                        + \overline{G}_{+1, \beta \gamma}^2
                    }}{2} t
                \right)
        \right] \right. \\ \left.
            \qquad \times \left[
                \left(a_0^{(-1)}\right)^* \left(a_0^{(\beta)}\right)^* \left(a_0^{(\gamma)}\right)^*
                e^{
                    i \frac{\sqrt{
                        \omega_1^2
                        + \overline{G}_{-1, \beta \gamma}^2
                    }}{2} t
                }
            - i A_{-1, \beta \gamma}^*
                \operatorname{sin}\left(
                    \frac{\sqrt{
                        \omega_1^2
                        + \overline{G}_{-1, \beta \gamma}^2
                    }}{2} t
                \right)
        \right] \right) \\
            \times \ketbra{+1_0}{-1_0} + \text{h.c.}
    \label{eq:Reduced Denwity Matrix 3}
\end{multline}

Just as our theorem had predicted, the presence of $\hat{H}_1$ and $\hat{H}_{\mathcal{EE}}$ in this Hamiltonian must be considered when looking at local dynamics as we can clearly see the presence of both these terms in the reduced density matrix in the form of $\omega_1$ and $\overline{G}_{\pm 1, \beta \gamma}$ within the arguments of the sine terms and complex exponentials.

We hypothesize that the influence of $\hat{H}_\mathcal{EE}$ on the local state stems from the generation of an \textit{emergent coupling} between the system and the second environment qubit, mediated by $\hat{H}_1$. This \textit{emergent coupling} would be akin to that found by Galapon \cite{Galapon_2016} in his quantum measurement model as he expanded his unitary time evolution operator using the Zassenhaus formula\footnote{Further discuttion on the Zassenhaus formula is found in Appendix \ref{ap:Zassenhaus}.}. Applying this expansion to the system's state vector
\begin{equation} \label{eq:State vector with emergent coupling}
    \ket{\Psi(t)}
        = e^{
            -i \left(
                \hat{H} - \hat{H}_1
            \right) t
            }
            e^{
            -i \hat{H}_1 t
            }
            U_{EC}(t) \ket{\Psi(0)},
\end{equation}
and comparing it to the state vector in terms of our chosen interaction representation reveals
\begin{equation}
    \ket{\Psi^{(\text{I})}(t)}
        = e^{-i \hat{H}_1 t}
            U_{EC}(t) \ket{\Psi(0)}.
\end{equation}
This shows that the time-evolution operator $U_{EC}(t)$ generated by the \textit{emergent coupling} is already encoded in the chosen interaction representation. This \textit{emergent coupling} would allow information to flow between the system and the second environment qubit, generating entanglement between these two subsystems.

To test this hypothesis, we analyze the entanglement between the local system and the second environment qubit. We choose concurrence as our measure of entanglement as it specifically measures the entanglement within a bipartite system, ignoring any entanglement with the environment. This allows us to measure entanglement using the bipartite reduced density matrix of these two subsystems $\hat{\rho}_{02}$. We use the description of concurrence introduced by Hill and Wooters \cite{Hill_1997, Wooters_1998},
\begin{equation}
    C(\hat{\rho}_{02}) = \operatorname{max} \{0, v_1 - v_2 - v_3 - v_4\},
\end{equation}
where $\{v_n\}$ are the square roots of the eigenvectors of the matrix $\hat{\rho}_{02} \tilde{\rho}_{02}$ arranged from greatest ($v_1$) to least ($v_4$), and $\tilde{\rho}_{02}$ is the spin-flipped state
\begin{equation}
    \tilde{\rho}_{02}
        = \left(
            \hat{\sigma}_0^y \otimes \hat{\sigma}_2^y
        \right)
        \hat{\rho}_{02}
        \left(
            \hat{\sigma}_0^y \otimes \hat{\sigma}_2^y
        \right).
\end{equation}

To get $\hat{\rho}_{02}$, we simply return to the density matrix of the entire system and partial trace over the degrees of freedom of only the first environment qubit $\{ \ket{\pm1_1} \}$,
\begin{multline}
    \hat{\rho}_{02}(t)
        = \sum_{\substack{\alpha, \alpha' \\ \gamma, \gamma '}}
            \sum_{\beta}
        \Biggl( \left[
            a_0^{(+1)} a_1^{(\beta)} a_2^{(\gamma)}
                e^{
                    -i \sqrt{
                        \omega_1^2
                        + \overline{G}_{+1, \beta \gamma}^2
                    } t/2
                }
            + i A_{+1, \beta \gamma}
                \operatorname{sin}\left(
                    \sqrt{
                        \omega_1^2
                        + \overline{G}_{+1, \beta \gamma}^2
                    } t/2
                \right)
        \right] \\
            \qquad \times \left[
                \left(a_0^{(-1)}\right)^* \left(a_0^{(\beta)}\right)^* \left(a_0^{(\gamma)}\right)^*
                e^{
                    i \sqrt{
                        \omega_1^2
                        + \overline{G}_{-1, \beta \gamma}^2
                    } t/2
                }
            - i A_{-1, \beta \gamma}^*
                \operatorname{sin}\left(
                    \sqrt{
                        \omega_1^2
                        + \overline{G}_{-1, \beta \gamma}^2
                    } t/2
                \right)
        \right] \Biggr) \\ \times
            e^{-i (\alpha - \alpha') \omega_0 t} 
            \ketbra{\alpha_0}{\alpha'_0}
                \otimes
            \ketbra{\gamma_0}{\gamma'_0}.
\end{multline}
This explicit equation for $\hat{\rho}_{02}$ will be the input for our numerical calculations using Matlab. Without loss of generality, we assume the initial probability amplitudes to be unbiased, i.e., $a_n^{(\eta)} = 1/\sqrt{2}$ for all $n$ and $\eta$.

\begin{figure}
    \centering
    \includegraphics[scale=1]{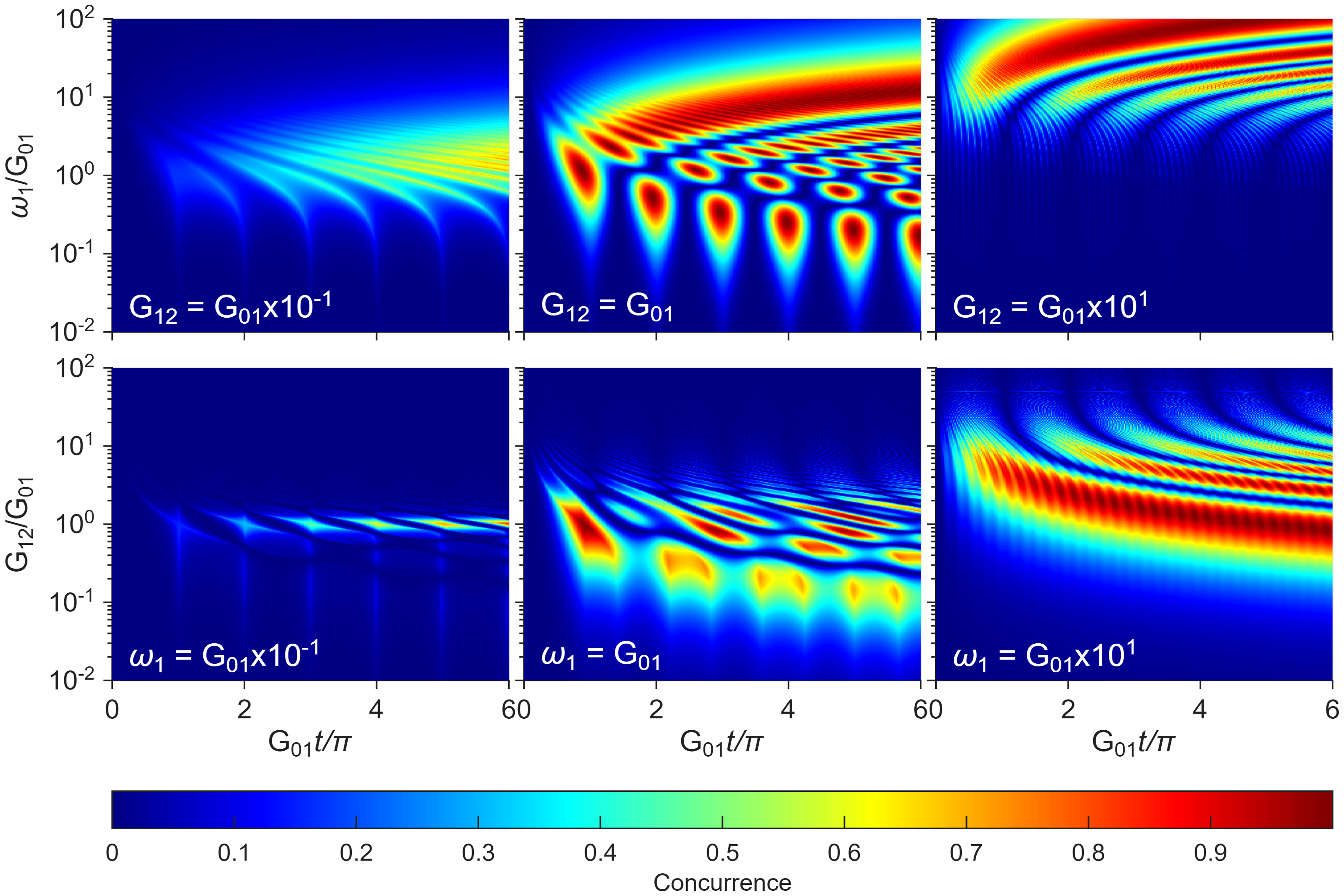}
    \caption{Time-evolution of concurrence under varying strength of transverse magnetic field for different intra-environment interaction strengths (top row) and vice versa (bottom row)}
    \label{fig:Concurrence}
\end{figure}

With this, we numerically analyze how the generation of concurrence is influenced by varying $\omega_1$ and $G_{12}$, generating the heatmaps in figure \ref{fig:Concurrence}. We choose not to explore the effects of varying $\omega_0$ on concurrence as Alporha et. al. \cite{Alporha_2025} have already shown that concurrence is independent on these magnetic fields. Thus, $\omega_0$ can be ignored without loss of generality. We also choose not to analyze the effects of varying $G_{01}$ as the bipartite system becomes symmetric with respect to the parity of the two subsystems after ignoring $\omega_0$. This symmetry means varying $G_{01}$ would only yield the same behavior as varying $G_{12}$.

In the top row of figure \ref{fig:Concurrence}, we set $G_{12}$ to be one order of magnitude weaker (left), equal (center), and one order of magnitude stronger (right) than $G_{01}$. We then analyze the concurrence dynamics as we vary $\omega_1$. From all three plots, we see that very little concurrence is generated for relatively weak $\omega_1$, regardless of the strengths of the interactions. This is to be expected from observing our earlier examples without $\omega_1$. In these examples, the degrees of freedom of the local system remain separable from those of the environment qubits not directly interacting with it. Thus, no entanglement is generated between these subsystems in the absence of $\omega_1$. We also see that, for strong entanglement to be created swiftly, $\omega_1$ should ideally be of a similar magnitude to the stronger of the two interactions. This observation is evidenced by the fact that the fastest rise in concurrence occurs around $\omega_1/G_{01} \sim 10^0$ when $G_{12}$ is weaker or equal $G_{01}$. Meanwhile, this fast rise in concurrence occurs at around $\omega_1/G_{01} \sim 10^1$ when we set $G_{12}=G_{01}\times10^1$. Lastly, entanglement seems to take longer to produce when when $\omega_1$ is significantly stronger than either of the two interactions.

We also analyze the concurrence dynamics for varying $G_{12}$ at different magnitudes of $\omega_1$ relative to $G_{01}$, shown in the bottom row of figure \ref{fig:Concurrence}. From all three plots, we see that weaker $G_{12}$ results in little to no concurrence. This is to be expected as, in the limit of $G_{12} \to 0$, the second environment qubit would be decoupled from the rest of the system. Thus, no information would flow to or from this qubit. We also see that there exists an upper bound to the relative strength of $G_{12}$ that can produce significant concurrence, where that upper bound is being dictated by $\omega_1$. Furthermore, we confirm our earlier analysis that little concurrence occurs when $\omega_1$ is relatively weak compared to the interactions, as shown in the first plot. We also confirm that a stronger $\omega_1$ results in slower generation of concurrence, requiring stronger $G_{12}$ to compensate but not exceeding the threshold set by $\omega_1$.

Regardless of the exact coherence dynamics of this system, these results show that entanglement can be generated between non-directly interacting subsystems so long as there is an \textit{emergent coupling}. We relate this \textit{emergent coupling} back to our theorem as its formulation from the Zassenhaus expansion requires a non-zero commutator, which would not satisfy our theorem's conditions.

\section{Conclusion}
In this paper, we have presented a theorem that identifies terms within any time-independent Hamiltonian that have no influence on local dynamics, local observables, or any local phenomena. Thus, these terms can altogether be ignored so long as only the local system is of concern. We substantiated this claim by providing two example Hamiltonians that can be simplified by this theorem. We also explored a system that our theorem could not simplify. We found that entanglement can be generated between the local system and the second environment despite a lack of direct interaction between the two. Though such a phenomenon has been described before \cite{Braun_2002, Qiu_2013, Dias_2023}, we were able to provide the mechanism by which such entanglement is generated, that being the presence of an \textit{emergent coupling}. Furthermore, we tie this exploration back to our theorem by arguing that the locally irrelevant terms identified by our theorem would also be incapable of generating an \textit{emergent coupling} towards the local system.

An interesting continuation of this study would be to further explore the entanglement dynamics of non-directly interacting qubits, especially for systems with an infinite-dimensional Hilbert space \cite{Galapon_2016}. Another possible extension would be to further probe the consequences of the \textit{emergent coupling} and its applications in quantum communication.

\appendix

\section{Partial Tracing} \label{ap:Partial Trace}

Given the density matrix written in terms of the orthonormal basis of the system $\ket{\psi_m}$ and of the environment $\ket{\phi_h}$,
\begin{equation} \label{eq:Density Matrix Appendix}
    \hat{\rho} (t) =
        \sum_{m, n, h, k} \rho_{m n h k} (t) \ket{\psi_m} \bra{\psi_n} \otimes \ket{\phi_h} \bra{\phi_k},
\end{equation}
describing a total compound system in the Hilbert space $\mathcal{H} = \mathcal{H}_a \otimes \mathcal{H}_b$, we can derive a reduced density matrix $\hat{\rho}_a (t)$ that describes only the characteristics of the subsystem in the subspace $\mathcal{H}_a$. This is done by tracing over the basis of subspace $b$
\begin{equation} \label{eq:Partial Tracing Appendix}
    \hat{\rho}_a (t)
    = \operatorname{Tr}_b \left[ \hat{\rho}(t) \right]
    = \sum_{m, n} \ket{\psi_m} \bra{\psi_h}
    \operatorname{Tr} \left[ \sum_{h, k} \rho_{mnhk} (t) \ket{\phi_h} \bra{\phi_k} \right].
\end{equation}
This process is also often dubbed as partial tracing \cite{Manzano_2020}.

For density matrices of separable states such as those encountered in the analyses of the first and second sample Hamiltonians, e.g.
\begin{equation} \label{eq:Separable Density Matrix Appendix}
    \hat{\rho} (t)
    = \left( \sum_{m, n} A_{m, n} (t) \ket{\psi_m} \bra{\psi_n} \right) \otimes \left( \sum_{h, k} B_{h, k} (t) \ket{\phi_h} \bra{\phi_k} \right),
\end{equation}
the unity trace property of density matrices makes it such that the reduced density matrix is trivially written as
\begin{equation} \label{eq:RDM of Separable Density Matrix}
    \hat{\rho}_a (t)
    = \sum_{m, n} A_{m, n} (t) \ket{\psi_m} \bra{\psi_n}.
\end{equation}

\section{Interaction Representation} \label{ap:Interaction Representation}

The interaction representation is an equally valid way of representing quantum systems along with the more commonly known Schr{\"o}dinger and Heisenberg representations. The interaction representation can be done for any Hamiltonian $\hat{H}$ consisting of two parts, the \textit{free part} $\hat{H}_0$ and the \textit{interaction part} $\hat{H}_1$. Any bifurcation of the Hamiltonian will yield a valid interaction representation. However, this representation is most useful when we select $\hat{H}_1$ as an additional term to a Hamiltonian with known solution $\hat{H}_0$. We apply this to the second and third sample Hamiltonians where part of the solutions are already demonstrated in the first example.

Assuming $\hat{H}$ is time-independent, as they are for all Hamiltonians discussed in this paper, operators $\hat{A} (t)$ in the interaction representation are given in terms of their Schr{\"o}dinger picture counterparts as $\hat{A}^{(\text{I})} (t) = e^{i \hat{H}_0 t} \hat{A} (t) \, e^{-i \hat{H}_0 t}$. Transformations of this form are solvable using the Baker-Hausdorff lemma\cite{Sakurai}. Meanwhile, state vectors are related to their Schr{\"o}dinger picture counterparts as $\ket{\Psi (t)} = e^{-i \hat{H}_0 t} \ket{\Psi^{(\text{I})} (t)}$, and evolve through time according to the equation of motion
\begin{equation} \label{eq:Interaction Representation State Vector equation of Motion Appendix}
    i \frac{\partial}{\partial t} \ket{\Psi^{(\text{I})} (t)} = \hat{H}_1^{(\text{I})} (t) \ket{\Psi^{(\text{I})} (t)},
\end{equation}
 where $\hat{H}_1^{(\text{I})} (t)$ is the interaction representation of $\hat{H}_1$ and often called the interaction potential \cite{Schwabl}.

Equivalently, the equation of motion  for the density matrix in the interaction picture is
\begin{equation} \label{eq:Interaction Representation Density Matrix Appendix}
    i \frac{\partial}{\partial t} \hat{\rho}^{(\text{I})} (t)
        = \left[
            \hat{H}_1^{(\text{I})} (t), \hat{\rho}^{(\text{I})} (t)
        \right],
\end{equation}
with the transformation $\hat{\rho} (t) = e^{-i \hat{H}_0 t} \hat{\rho}^{(\text{I})} (t) e^{i \hat{H}_0 t}$ to convert the density matrix from the interaction picture to the Schr{\"o}dinger picture. Here, we can solve for the state vector in the interaction picture for both the second and third sample Hamiltonians.

Starting with the second sample Hamiltonian we assume the state vector in the interaction picture has the form
\begin{equation} \label{eq:Interaction State Vector 2 Form}
    \ket{\Psi^{(\text{I})} (t)}
        = \bigotimes\limits_{n=0}^{N-3}
            \left(
                a_n^{(+1)} \ket{+1_n} + a_n^{(-1)} \ket{-1_n}
            \right)
            \otimes \sum_{\alpha, \beta, \gamma}
                C_{\alpha \beta \gamma} (t)
                    \ket{\alpha_{N-2}}
                    \ket{\beta_{N-1}}
                    \ket{\gamma_{N}},
\end{equation}
with $\alpha, \beta, \gamma \in \{-1, +1\}$. Having time-evolution occur only in the probability amplitudes of the states of the last three qubits is reasonable as the interaction potential shown in equation \eqref{eq:Interaction Potential 2 Solved} consists of operators that act only on the degrees of freedom of these qubits. As a result, time evolution is only generated for these degrees of freedom and the degrees of freedom of all other qubits remain stationary.

Substituting equations \eqref{eq:Interaction Potential 2 Solved} and \eqref{eq:Interaction State Vector 2 Form} to equation \eqref{eq:Interaction Representation State Vector equation of Motion Appendix} yields the differential equation
\begin{multline} \label{eq:equations of Motion 2}
    \sum_{\alpha, \beta, \gamma}
        i \frac{\partial}{\partial t} C_{\alpha, \beta, \gamma}
            \ket{\alpha_{N-2}}
            \ket{\beta_{N-1}}
            \ket{\gamma_{N}} \\
    = \sum_{\alpha, \beta, \gamma}
        \frac{1}{2} i (\gamma - \beta) D
            e^{
                i \alpha \beta
                    G_{N-2, N-1} t
            }
            C_{\alpha, \beta, \gamma}
                \ket{\alpha_{N-2}}
                \ket{- \beta_{N-1}}
                \ket{- \gamma_{N}}.
\end{multline}
For the arbitrary state $\ket{\alpha_{N-2}} \ket{\beta_{N-1}} \ket{\gamma_{N}}$, the differential equation becomes
\begin{equation} \label{eq:ODE 2 Appendix}
    i \frac{\partial}{\partial t} C_{\alpha, \beta, \gamma} (t)
        = \frac{i}{2} (\beta - \gamma) D e^{i \alpha \beta G_{N-2, N-1} t} C_{\alpha, -\beta, -\gamma} (t),
\end{equation}
which yields the state vector shown in equation \eqref{eq:Interaction State Vector 2}. The coefficients given by equation \eqref{eq:Coefficient 2} arise from imposing the initial conditions from equation \eqref{eq:Initial State 2} to this solution and its differential equation.

Similarly for the third sample Hamiltonian, we get the differential equation
\begin{equation}
    i \frac{\partial}{\partial t} C_{\alpha, \beta, \gamma} (t)
        = \frac{1}{2} \omega_1
            e^{i\beta (\alpha G_{01} + \gamma G_{12})t}
            C_{\alpha, -\beta, \gamma} (t)
\end{equation}
for the time-evolving probability amplitudes $C_{\alpha, \beta, \gamma} (t)$ of the state $\ket{\alpha_{0}} \ket{\beta_{1}} \ket{\gamma_{2}}$ in the interaction picture, yielding equation \eqref{eq:Interaction State Vector 3}. Again, the coefficient given by equation \eqref{eq:Coefficient 3} arises from applying the initial conditions in equation \eqref{eq:Initial State 3} to this solution and its differential equation.

\section{Zassenhaus Expansion} \label{ap:Zassenhaus}
The Zassenhaus formula states that we can expand expressions of the form $e^{X+Y}$ for non-commuting variables $X$ and $Y$ as
\begin{equation} \label{eq:Zassenhaus Appendix}
    e^{X+Y} = e^X e^Y \prod_{n=2}^\infty e^{C_n(X, Y)},
\end{equation}
where $C_n(X, Y)$ are polynomials in X and Y of degree $n$. In terms of the nested commutators of $X$ and $Y$, the first few terms of this series are
\begin{align}
    C_2(X, Y) &= -\frac{1}{2}[X, Y], \label{eq:Zassenhaus C2}\\
    C_3(X, Y) &= \frac{1}{3}[Y, [X, Y]] + \frac{1}{6}[X, [X, Y]], \label{eq:Zassenhaus C3}
\end{align}
and so on \cite{Casas_2012}.

This formula allows us to expand the unitary time-evolution operator for time-independent Hamiltonians. This separates them into individual complex exponentials and explicitly provides the unitary time-evolution operator generated by the \textit{emergent coupling} via the expansion's higher order terms in the $\Pi$ product, as shown in equation \eqref{eq:State vector with emergent coupling}.

\bibliography{References}

\end{document}